\documentclass[
showkeys,
twocolumn,
superscriptaddress,
showpacs,
 amsmath,amssymb,
 aps,
pre,
]{revtex4-1}
\usepackage{natbib}
\usepackage{graphicx}
\usepackage{dcolumn}
\usepackage{bm}

\begin{document}

\title{Dynamic Phase Transition in the Contact Process with Spatial 
Disorder: Griffiths Phase and Complex Persistence Exponents}
\author{Priyanka D. Bhoyar}
 \affiliation{Department of Physics,
 Seth Kesarimal Porwal College of
Arts and Science and Commerce, Kamptee, 441 001, 
 India.}

\author{Prashant M. Gade }%
\email{prashant.m.gade@gmail.com}
\affiliation{%
 Department of Physics, 
Rashtrasant Tukadoji Maharaj Nagpur 
University, Nagpur, 440 033, India.
}%
\date{\today}

\begin{abstract}

We present a model which displays Griffiths phase 
{\it{i.e.}} algebraic decay of density 
with continuously varying exponent 
in the absorbing phase. In  active
phase, the memory of initial conditions is lost with continuously
varying complex exponent in this model.
This is 1-D model where fraction $r$ of sites
obey rules leading to directed percolation (DP) class
and the 
rest evolve according to rules leading to compact directed percolation
(CDP) class.
For infection probability $p < p_c$, the fraction
of active sites $\rho(t)=0$ asymptotically. For $p>p_c$, 
$\rho(\infty)>0$.
At $p=p_c$, $\rho(t)$, the survival probability from single seed  and 
the average number of active sites starting from single seed
decay logarithmically.
The local persistence $P_l(\infty)>0$ for $p<p_c$ and $P_l(\infty)=0$ for
$p>p_c$.
For $p>p_s$, local persistence $P_l(t)$ decays as a power law with  
continuously varying exponents. The persistence exponent is clearly complex as 
$p\rightarrow 1$. The complex exponent implies
logarithmic periodic oscillations in persistence.
The wavelength and
the amplitude of the logarithmic periodic oscillations increases with $p$.
We note that underlying lattice or disorder does not have self-similar 
structure.
\end{abstract}


\maketitle
\section{Introduction}
Understanding of phase transition in equilibrium
statistical physics
is a major success in theoretical physics in the '70s. 
The concepts of scaling and renormalization group were
introduced to explain the divergence of characteristic
length-scales near continuous phase transitions.
Extension of these ideas to 
the nonequilibrium systems 
is an active area
of research with applications ranging from
granular matter to epidemics \cite{marro2005nonequilibrium,
domb1995statistical,hinrichsen2000non}.
The power-laws 
associated with the continuous transitions are observed
in very close vicinity of the critical point.
They are found precisely in theoretical models and
are a signature of the universality class of associated phase
transition. Very few experimental verifications
of such power-laws are obtained. 
It requires extremely fine-tuning of experimental parameters.
Nonetheless, the power-laws in space and time
are ubiquitous. This has led to the paradigm
of self-organized criticality which often
requires adiabatic drive \cite{bak1987self,pruessner2012self,carlson1993self}. 
Another reason for power-laws 
can be disorder and inhomogeneities \cite{noest1986new,noest1988power,
webman1998dynamical}.

Usually, the power-laws have real exponents. 
A complex exponent will
lead to log-periodic oscillatory corrections to the power-law.
Such oscillations have
been obtained or predicted in a few 
situations \cite{sornette1998discrete,akkermans2012spatial,saadatfar2002diffusion}. 
For example, 
observation of log-periodic oscillations in the stock market has been
associated with the possibility of a crash.
Some understanding
of such oscillations at the critical point when underlying
lattice or disorder has self-similar characteristics is obtained
\cite{johansen1999predicting,sornette1998discrete,barghathi2014contact}. In this work,
we observe
log-periodic oscillations in the memory of initial conditions
even when there is no self-similar structure in the lattice or
disorder and the system is far away from the critical point.
This can be interpreted as an outcome of quenched disorder alone.

Most theoretical studies in the equilibrium and the nonequilibrium phase 
transitions
involve idealized homogeneous systems. It is a useful approximation
(like equilibrium). But the real-life
systems involve
inhomogeneities invariably. 
They play an important
role in several experiments. For example, in the catalytic reactions,
the catalytic
surface is not homogeneous. The inhomogeneities
can change or destroy critical behavior \cite{webman1998dynamical}.
The theory of
spin-glasses in equilibrium systems
has given useful insights in disordered systems. It
has found applications in fields ranging from evolution to
computer science \cite{fischer1993spin}.

In this work, we focus on the absorbing phase transitions in nonequilibrium
processes in the presence of spatial disorder. We take prototypical and
widely studied class of directed percolation (DP).
It is characterized
by one component order parameter, short-range dynamical rule,
no additional symmetries and no
quenched disorder \cite{janssen1981nonequilibrium,grassberger1982p}.
The experimental verification of DP is rare, although it is
very well studied in theory and simulations. The experimental
verifications  are related to
the spatiotemporal
intermittency \cite{hinrichsen2000possible}.
There are some more universality classes for the transition
to an absorbing phase.
One of them is
compact directed percolation (CDP).
In this class, if all the neighbors of a given site are
active, it becomes active with probability one \cite{henkel2008non}.

Certain justifications have been proposed for
difficulties in observing DP behavior
in experiments \cite{hinrichsen2000possible}.
Intrinsic fluctuations
may smear out the transition to the truly absorbing phase.
DP does not take into
account possible intermediate phases in
the reaction sequence.  
The nature of the update, (parallel or random sequential),
can change the nature of the transition.
Finally, realistic systems cannot avoid random inhomogeneity of
some kind. Not only the
experimental values of the exponents 
may vary from their theoretical prediction,
but the critical behavior may be destroyed due
to the inhomogeneity \cite{vojta2006rare}.

We study the impact of spatially quenched disorder on DP
universality class. We consider a very
strong disorder in which the universality class of the underlying
system changes completely.
Several studies have been carried out to study the effect of
spatially quenched disorder in the past. However, a clear picture
is still eluding. The studies on the effect of quenched disorder
on the DP model show that even very weak randomness can drastically
modify the phase diagram and critical behavior \cite{webman1998dynamical}.
According to Harris criteria, the quenched disorder are relevant
perturbations if $\textit{d} \nu_{\perp}<2$ where $\textit{d}$,
is the dimensionality and $\nu_{\perp}$ is the correlation
length exponent in the spatial direction of the pure
system \cite{harris1974effect}. In particular, we consider 
disorder in which half the sites obey DP and the other half obey
CDP rules. This is a very strong perturbation. 
For $d=1$, $\nu_{\perp}=1.09$ for DP and $\nu_{\perp}=1$ for CDP. 
Since, $\nu_{\perp}<2$ in either case, we can expect quenched
disorder to be a relevant perturbation. 

In CDP universality class, 
we observe compact percolation clusters.
It is characterized by an additional $\mathbb{Z}_2$ symmetry. Here, 
an active site with all active neighbors cannot become
inactive.
 The transition is
governed by random walks at the end of the string \cite{henkel2008non}.

We find that the presence of random spatial inhomogeneity leads to change
in the critical behavior and ultra-slow dynamics in the 1-D model.
The system undergoes algebraic decay
with changing exponents in a part of the absorbing phase, followed
by stretched exponential and exponential decay. The region with
generic power-laws is known as Griffiths phase.
It is known to emerge from rare
region effects due to the presence of
quenched defects \cite{griffiths1969nonanalytic}.
The rare regions locally favour one phase
over the other \textit{i.e.} although the bulk is globally in the absorbing
phase, the rare regions are locally in the fluctuating phase with over
average percolation probability. The dynamics of
rare regions are extremely slow which leads to stretched exponential decay
or algebraic decay \cite{vojta2003disorder}.

Of late, there has also been interest in the possibility of determining
the nature of phase transition from short-time dynamics. The short-time dynamics
of the spin-glass model and the Baxter-Wu model were studied.
Novel
critical exponents unrelated to known exponents were
obtained \cite{arashiro2003short,huse1989remanent}.
Other quantifiers such as persistence exponents have
also been studied. 
We studied local persistence
for this study which quantifies the loss of memory
of the initial conditions. In general,
power-law decay of the persistence and well-defined persistence
exponents (if any) are obtained only at the critical point.
However, for our model,
we observe well-defined persistence exponents
over the entire fluctuating phase. These exponents are complex and we observe
logarithmic periodic oscillations over and above the usual power-laws. The
oscillations do not average out (by the cancellation of phase)
by averaging over disorder and initial conditions. Such
exponents have not been reported before to the best of our knowledge.

\section{Model}
 We consider the cellular automata model of contact process (CP)
originally proposed by Domany and Kinzel \cite{domany1984equivalence,
kinzel1985phase}. This model
shows the transition in DP or CDP class.
We consider a 1-D lattice of length $N$.
Each site $i$ is associated with variable
$v_i(t)$ which is 0 or 1 depending whether they are 'inactive' or 'active'.
Each site at any time t+1 becomes active with 
certain probability depending on  its neighbors' state
at the previous time t.
The conditional probabilities
$P(\textit{v}_i(t+1))|\textit{v}_{i-1}(t),\textit{v}_{i+1}(t))$ are
defined as follows:
P(1$|$0,0)=0,  P(1$|$1,1)=q,  P(1$|$1,0)=P(1$|$0,1)=$p$.
DP transition can be obtained for 
$q \ne 1$. 
Let us consider $p=q<1$ for simplicity. 
The order parameter, the fraction of active sites, 
is given by $\rho(t)=\frac{1}{N}\sum_{i=1}^N v_i(t)$.
Below the critical
probability $p_c$, the cluster goes
to an absorbing phase from which it cannot escape. Above $p_c$,
$\rho(t)>0$.
In this case
 $\rho(t) \sim exp(-\lambda t)$ for $p<p_c$ 
after a brief transient and
$\rho(\infty)>0$
 asymptotically for $p>p_c$.
At $p=p_c$, $\rho(t)$ decays as a power
law $\rho(t) \sim t^{-\delta}$ with $\delta=0.158$ in 1-D. For the above model,
$p_c=0.705$.

For CDP $q=1$ and $p<1$.
Thus, no inactive sites can be created in a continguous region of
active sites. By symmetry, $p_c=0.5$ and $\rho(\infty)$ 
jumps from 0 to 1 at $p_c$.
The exponent $\delta=0$ for CDP.

Now, we consider a model in which a fraction of sites $r$ 
marked as type A evolve
according to CDP rules and the remaining $1-r$ fraction of sites
marked as type B 
evolve according to
DP rules. 
For the contact process modelling diseases, the type A particles can be
interpreted as children, elderly or sick people who are extremely vulnerable
and will catch a disease if everyone around them is sick.
The rules are symmetric and our update is synchronous.
The conditional probabilities of the update are mentioned above.
For the particles of type A, $q=1$, 
and for type B, $q=p, p<1$. We consider the model for $r=0.5$.

We compute two quantities (a) the fraction of active sites $\rho(t)$, and
(b) the local persistence $P_l(t)$. $P_l(t)$ is the fraction of
sites such that
$v_i(t)=v_i(t')$ for $0\le t'\le t$.  These  sites have
not changed their state {\it{even once}} from their
initial conditions till time $t$. This is a non-Markovian
quantity.
Interestingly, it has been found that
this quantity displays a power-law decay at the critical point
of dynamic phase transition in some cases. 
In these cases,
$P_l(t)\sim 1/t^{\theta_l}$ at the 
critical point where $\theta_l$ is known as the local 
persistence exponent.  This is a new exponent independent
from other critical exponents related to the transition. 
It is not universal. However, it has been found useful
in finding other exponents such as $z$ and $\nu_{\parallel}$ \cite{
fuchs2008local,menon2003persistence}.

The quantity $\rho(t)$ is an
order parameter for the absorbing state transitions while $P_l(t)$ is
an order parameter for the spreading transition. 
$\rho(\infty)=0$ implies that there are no active sites in the
lattice asymptotically and evolution has effectively stopped.  
$P_l(\infty) \ne 0$  suggests that some sites
do not deviate from their
initial conditions even once ever. The persistent sites partition
lattice in various
clusters such that there is no spread of information from one cluster
to another. 
Both $\rho(\infty)$ as well as $P_l(\infty)$ can give us information 
about the phase transition. We average over initial
conditions as well as disorder realizations.

The order parameter 
$\rho(\infty)>0$ for $p >p_c$.
We find that for $r=0.5$, $p_c=0.651$ and
a finite density of active sites is obtained for $p>p_c$.
Normally,  all inactive
sites become active at some point of time in the active phase and
$P_l(\infty)$
asymptotically approaches zero for $p>p_c$.
On the other hand, $P_l(\infty)>0$ for $p<p_c$. 
However, it will be shown in next section that at $p=p_c$, the
active sites decay logarithmically and dynamics is extremely 
slow and we do not obtain power-law decay of
persistence.
The persistence decays as power law for $p>p_s$.
For $p_c<p<p_s$, the decay
is slower than 
power law.
The persistence exponent is the largest at $p=p_s$ and 
reduces as $p \rightarrow 1$

\begin{figure}[hbt!]
\scalebox{0.3}{
	\includegraphics{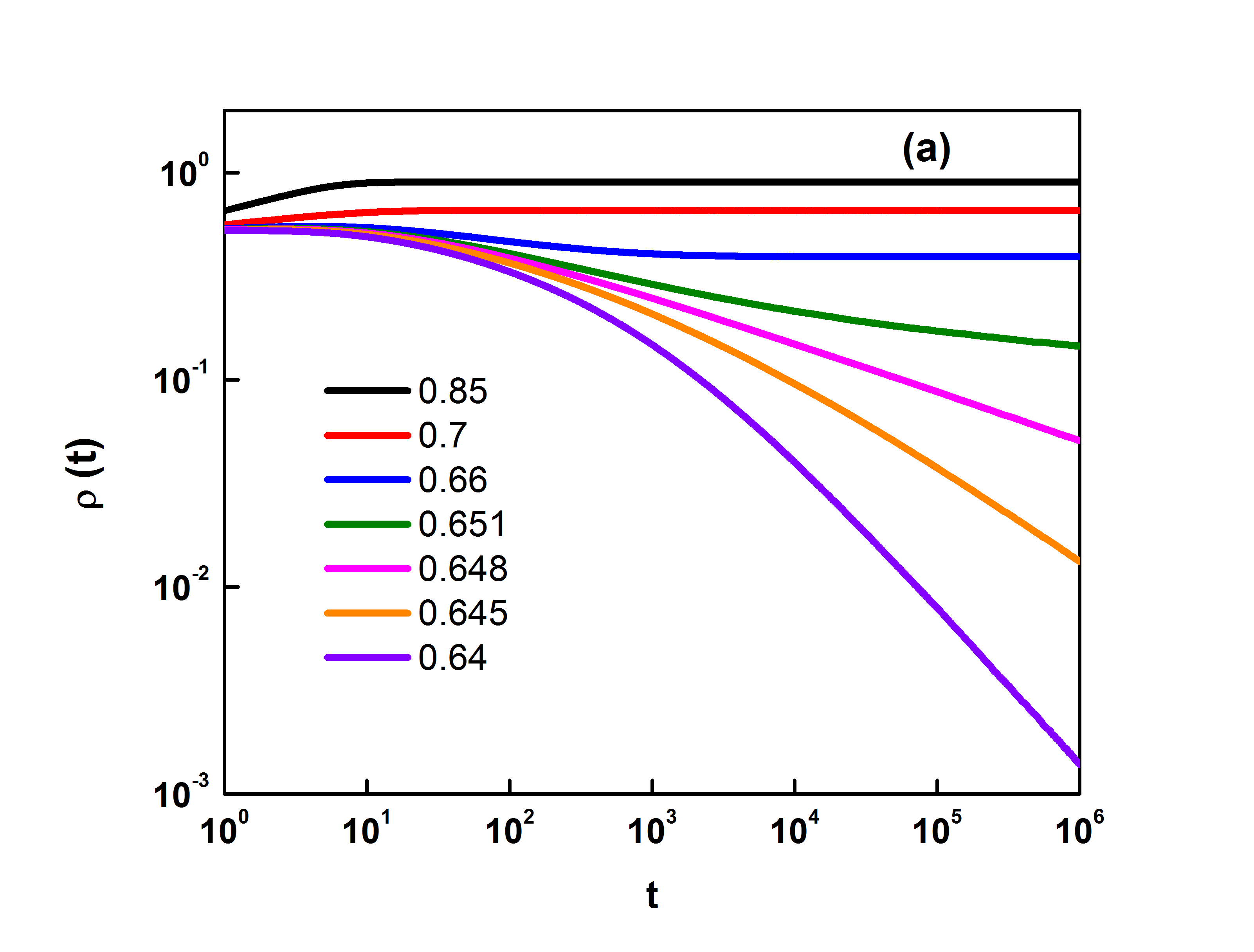}
}
\scalebox{0.3}{
      \includegraphics{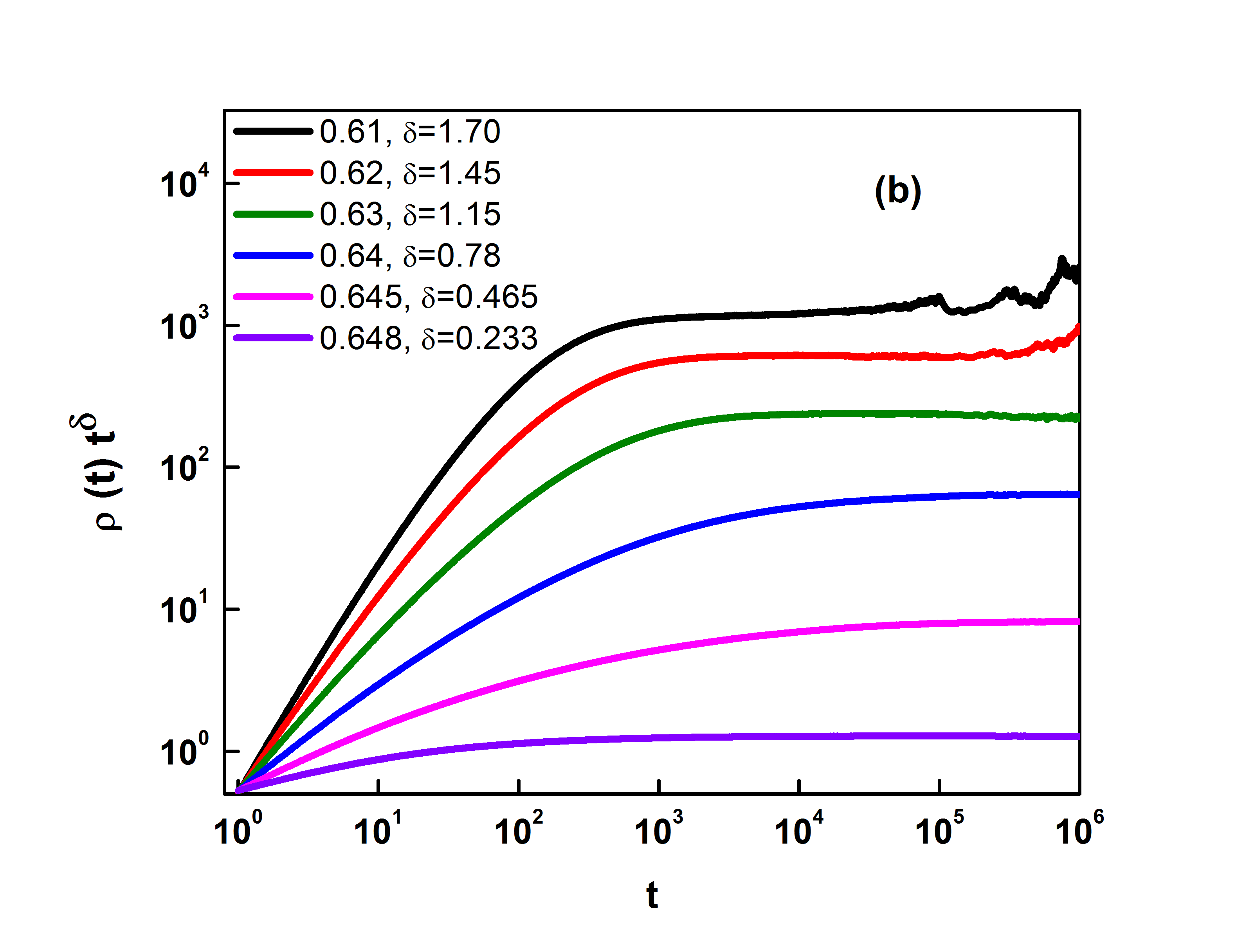}
}
	    \caption{\label{fig:1}(a)Overview of time evolution of
	average density $\rho(t)$ of 1-D system of size $N=2.5\time 10^6$
	for $r=0.5$ and $p\ge p_c$, where $p_c=0.651$. We find that
	$\rho(t)\sim t^{-{\delta}}$
	for $p$ is close to $p_c$ and $p<p_c$.
	(b) Log-log plot of $\rho(t) \times t^{\delta}$ $\it{vs}$ time
	in the Griffiths region for system with $p<p_c$.
	Exponent $\delta$ changes continuously and reaches $0$
	as $p\rightarrow p_c$.} 
\end{figure}

\begin{figure}[hbt!]
\scalebox{0.3}{
    \includegraphics{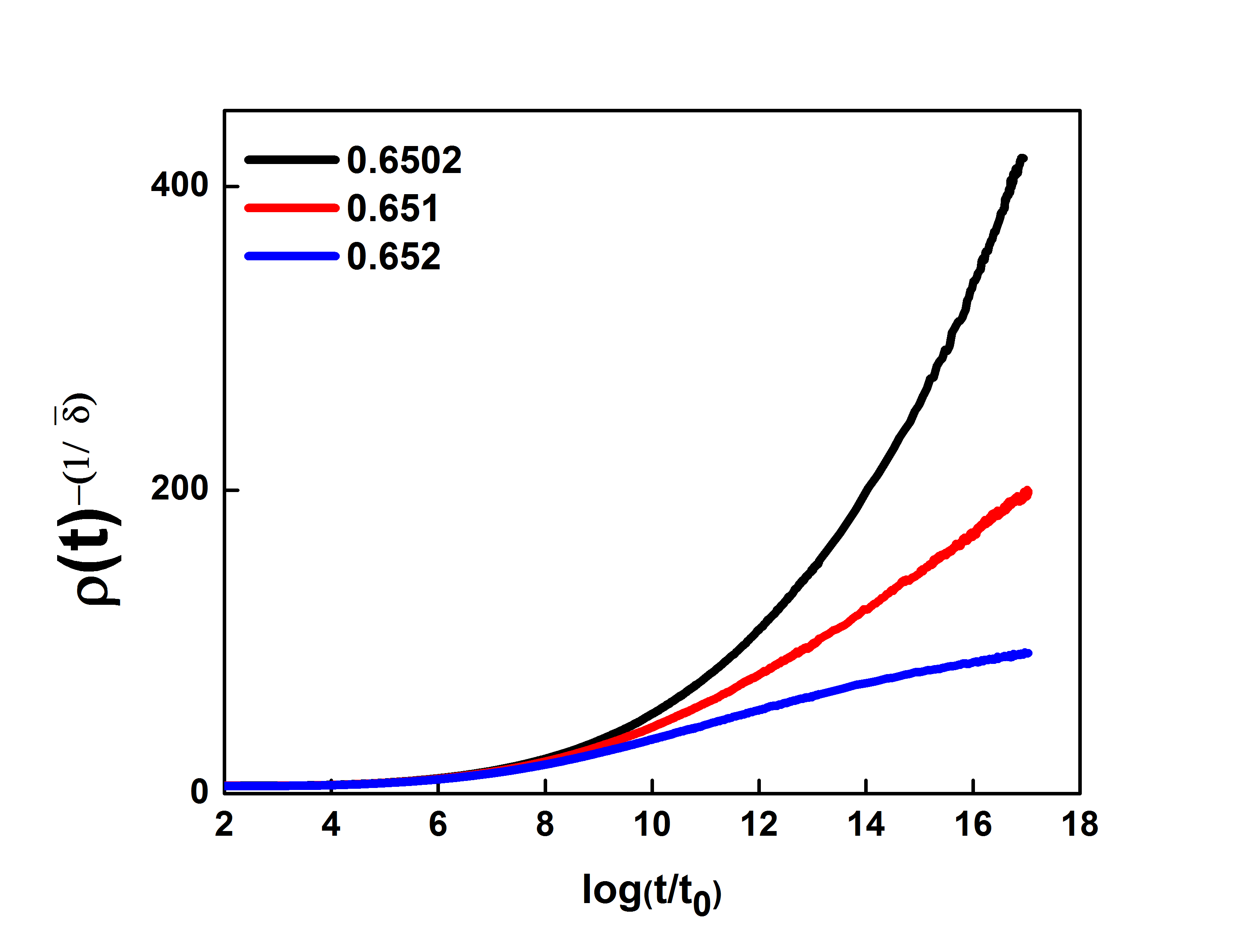}
}
    \caption{The time evolution of density of active sites $\rho(t)$
	for a system of size $2.5 \times 10^6$ with $p<p_c$, $p=p_c$
	and $p>p_c$. $\rho(t)$ decay as $[\log(t/t_0)]^{-\bar{\delta}}$
	where $\bar{\delta}=0.381$. only at $p=p_c$.}
\label{fig:2}
\end{figure}

\section{Simulation and Results}
We simulate the system for
$2.5 \times 10^6  $ sites and average more than  $10^3$ 
configurations.

(a) The absorbing state transition:
For $p<p_c=0.651$, all sites become
eventually inactive and the evolution stops. Normally, we observe
an exponential decay of $\rho(t)$ for $p<p_c$ and 
a power-law decay 
is observed at $p=p_c$.  
However, in our case,
a power-law decay of $\rho(t)$ is observed over a range of
parameters below the critical point but
not at the critical point, {\it {i.e.}} $\rho(t)
\sim 1/t^{\delta}$ with decreasing $\delta$, as $p \rightarrow p_c$.
The regime in which
the power-law decay of $\rho(t)$ with changing exponents is observed is
known as the Griffiths phase. 
The relaxation behavior changes to stretched
exponential and eventually to exponential for very small values of $p<<p_c$.
Fig.\ref{fig:1}(a) shows the density
of active sites $\rho(t)$  $\textit{vs}$ time
$t$ on logarithmic scale for various values of $p$.
Relaxation is slower
as $p \rightarrow p_c$ and $\delta \rightarrow 0$.
The power-law decay $\rho(t)\sim t^{-\delta}$ implies that 
$\rho(t) \times t^{\delta}$
approaches a constant value asymptotically.
This expectation is indeed fulfilled.
Fig.\ref{fig:1}(b) shows
 $\rho(t) \times t^\delta$ with 
 $t$ on logarithmic scale for $p<p_c$, close to $p_c$.
As $p \rightarrow p_c$, $\delta$ decreases. 

This transition has
been confirmed by large lattice simulation mentioned 
above as well as by simulations starting with a single seed.
At $p=p_c$ the relaxation is ultra-slow. For a large lattice
of size $2.5\times 10^6$, we observe logarithmic decay of $\rho(t)$
at $p=p_c$.
The transition is expected to be in the universality class of activated
scaling \cite{vojta2005critical,vojta2009infinite,hooyberghs2004absorbing,hooyberghs2003strong}.
For this class, the 
the proposed behavior is $\rho(t) \sim [\log(t/t_0)]^{-\bar{\delta}}$.
Thus, if we plot $\rho(t)^{-{\frac{1}{\bar{\delta}}}}$ as 
a function of $\log(t/t_0)$ we observe linear
behavior only at the critical point. 
In  Fig.\ref{fig:2}, we have plotted 
$\rho(t)^{-\frac{1}{\bar{\delta}}}$ as 
a function of $\log(t/t_0)$ Fig
with $\bar{\delta}=0.381$
at $p<p_c$, $p=p_c$ and $p>p_c$. The value
 $\bar{\delta} \sim 0.381$ is close to previously obtained
value $\bar{\delta}=0.38197$.  
   
\begin{figure}
\scalebox{0.3}{
               \includegraphics{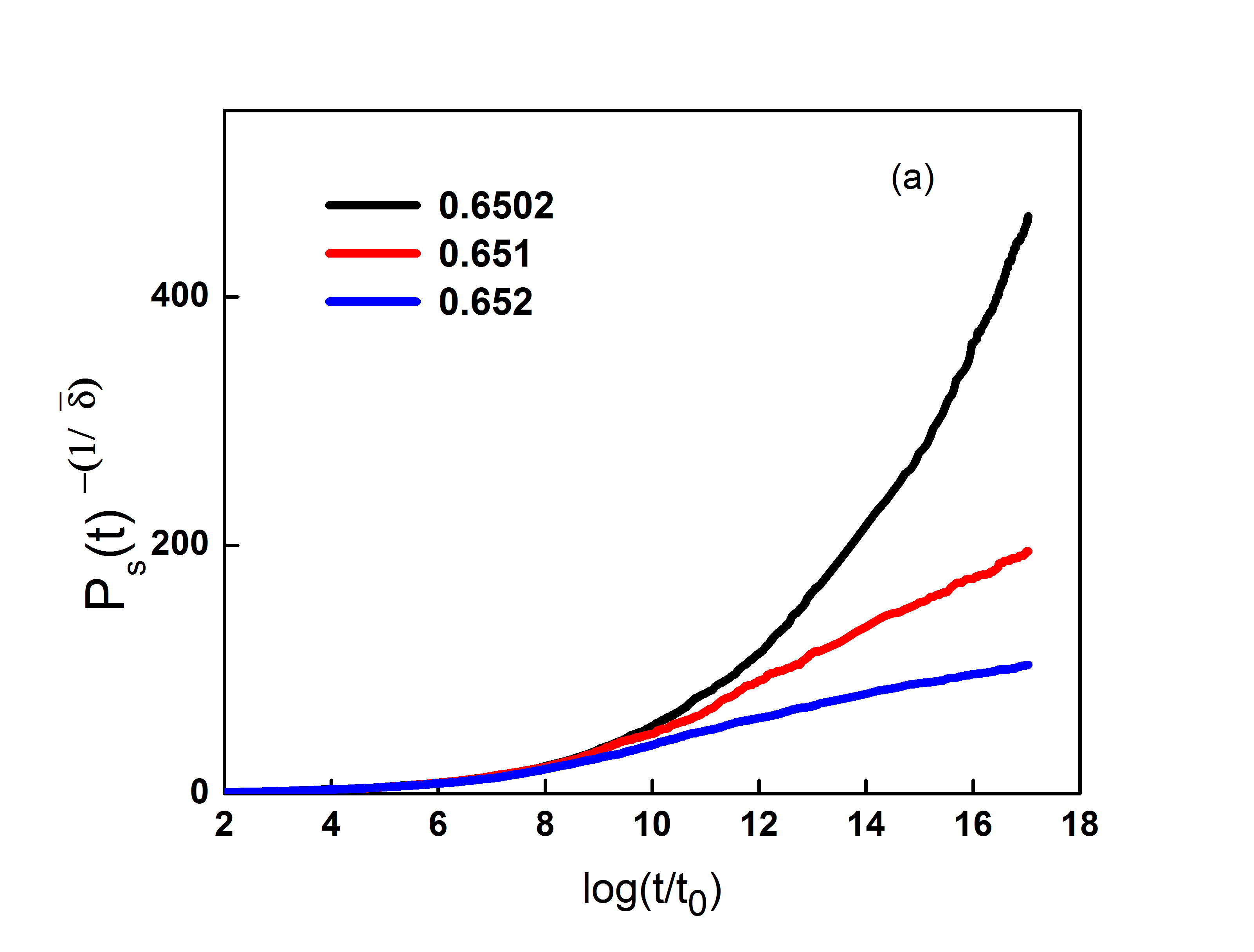}
}
\scalebox{0.3}{
               \includegraphics{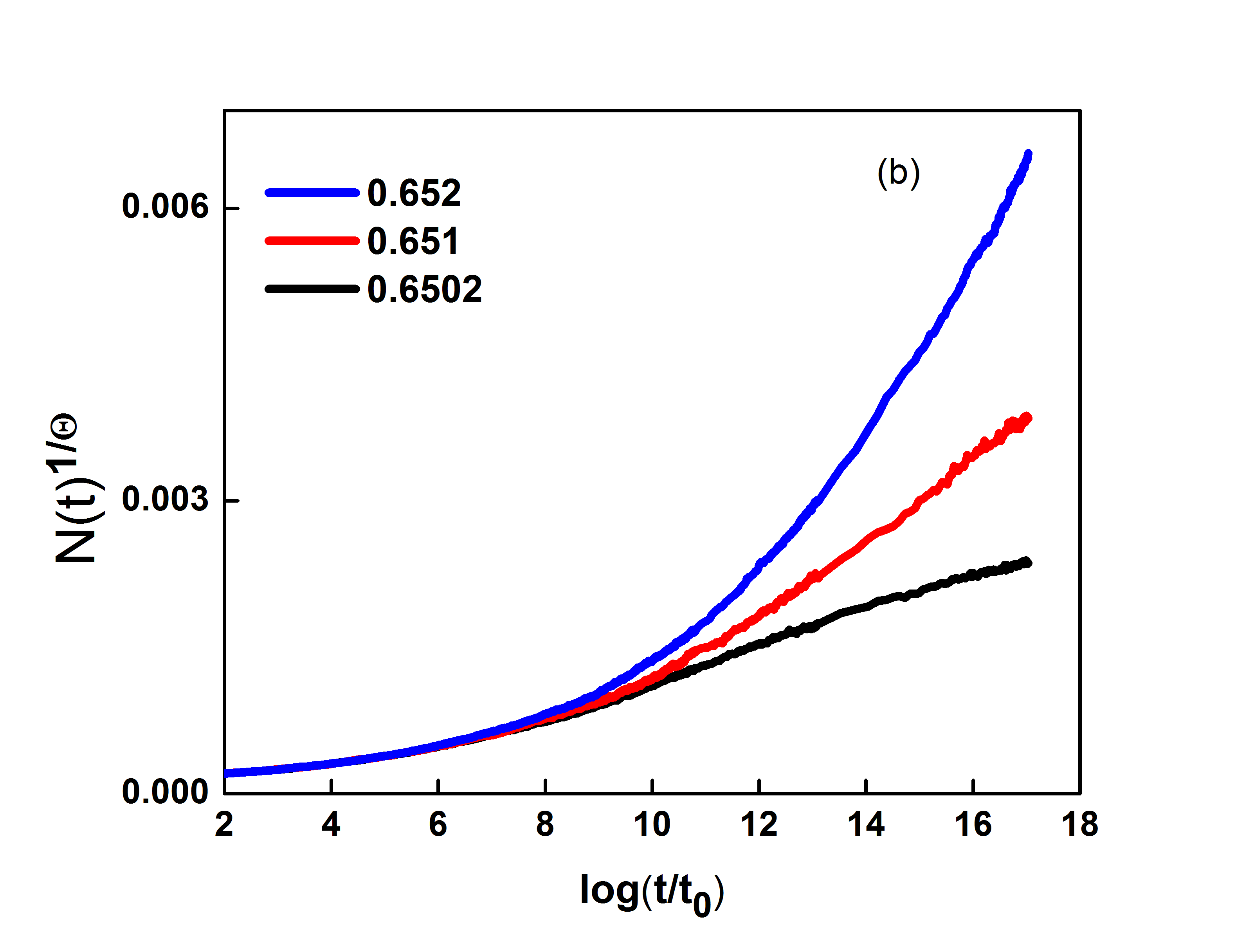}
}
   \caption{(a)The time evolution of the survival probability $P_s(t)$
	for disorder concentration $r=0.5$ starting with a single seed,
	for $p<p_c$, $p=p_c$ and $p>p_c$.
	$P_s(t)$ decay as $[\log (t/t_0)]^{-\bar{\delta}}$
	with $\bar{\delta}=0.381$.
	(b) The time evolution of average number of active sites in a cluster
	starting with a single seed $N(t)$.$N(t)$ decay as
	$[\log (t/t_0)]^{\Theta}$ with $\Theta=1.236$ and $t_0=0.2$.}
\label{fig:3}
\end{figure}

   Due to logarithmic decay, it is extremely difficult to locate 
   the critical point precisely and further tests are required to
   locate it.
   We confirm the above value with single seed simulations. 
   We compute (a)the survival probability (fraction of
clusters surviving till time $t$) $P_s(t)$ and (b) the average number of
particle in a cluster starting from single seed $N(t)$. 
At $p=p_c$, (a) the quantity $P_s(t)$ is expected to decay
asymptotically as $P_s(t) \sim [\log (t/t_0)]^{-\bar{\delta}}$ and
(b) $N(t)$ decays as $N(t) \sim [\log (t/t_0)]^{\Theta}$.
Again we plot $P_s(t)^{-{\frac{1}{\bar{\delta}}}}$ as a 
function of $\log(t/t_0)$ and obtain
linear behavior only for $p=p_c$
(See Fig.\ref{fig:3}(a)).
Similarly, if we plot $N(t)^{\frac{1}{\Theta}}$ as a function of
$\log(t/t_0)$, a linear behavior is expected only at $p=p_c$.
We observe that for $t_0=0.2$, $\bar{\delta}=0.381$ and 
$\Theta=1.236$, linear behavior is obtained only
at $p=0.651$. 
(See Fig.\ref{fig:3}(b)).
The values of $\bar{\delta}$ and $\Theta$ match with those
expected in the class of activated scaling \cite{hooyberghs2004absorbing,hooyberghs2003strong}.
Thus, we confirm $p_c=0.651\pm 0.0005$ in three
different ways using both large lattice simulations as well as single
seed simulations. 

\begin{figure}[hbt!]
\scalebox{0.3}{
    \includegraphics{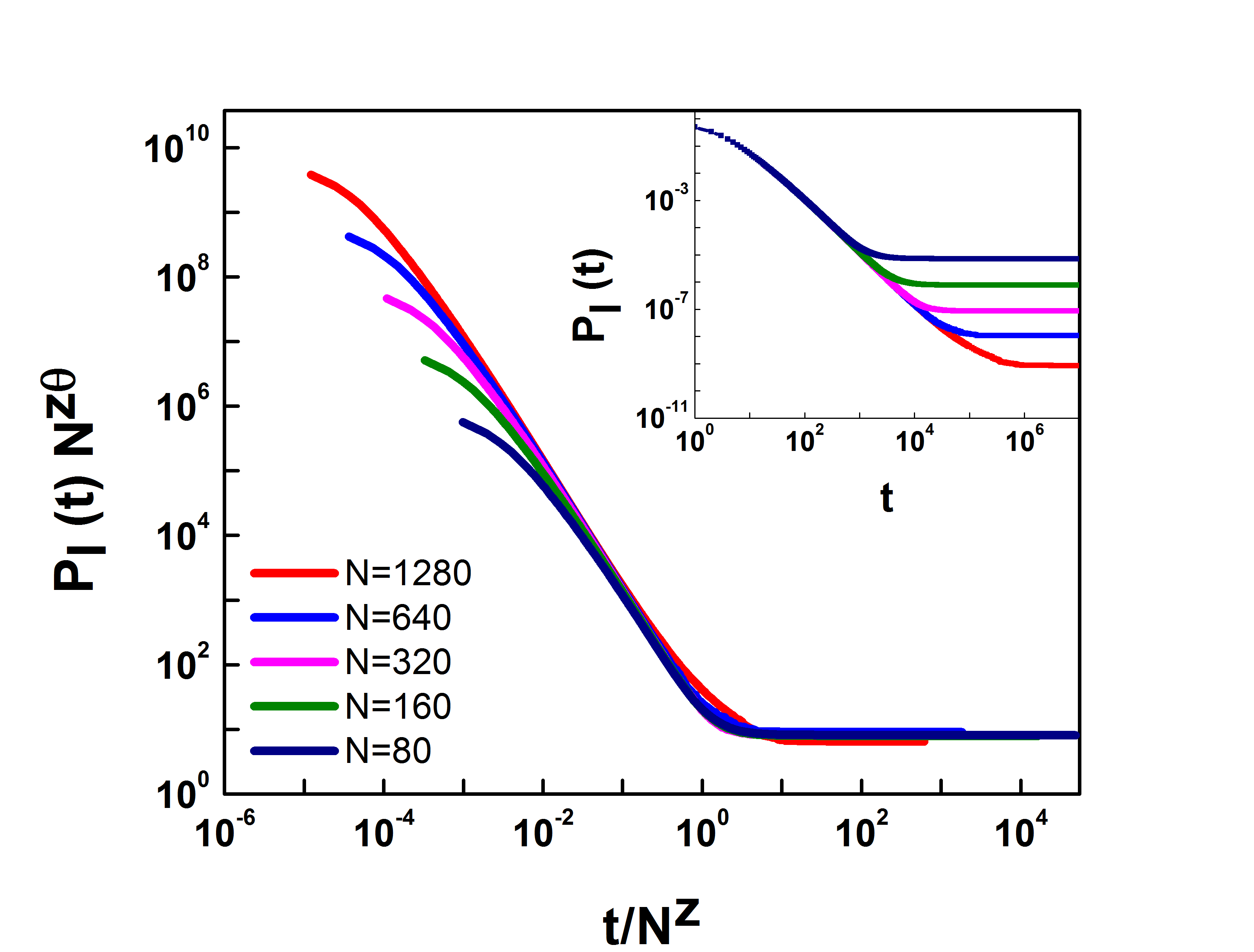}
}
	\caption{ Scaling plot of the local Persistence $P_s(t)$ at $r=0.5$.
	with $p=p_s=0.655$ for different size of lattice N. $P_s(t)$
	is scaled with $N^{z\theta}$.
	Inset:Time evolution of
	$P_l(\infty)$ with $r=0.5$, $p=p_s$ for various
	sizes of lattice N=80,160,320,640,1280}
\label{fig:4}
\end{figure}

(b) Persistence: We compute $P_l(t)$ as a function of time
for a large system of size $N=2.5\times 10^6$.  
Usually, the absorbing state 
transition is accompanied by the spreading transition. (In
the active phase, we expect
the inactive sites can be expected to become active 
at some time and the active sites will become inactive due to 
fluctuation.) 
Since the active sites decay logarithmically
at $p=p_c$, the persistence decays very slowly at the
critical point and cannot be fitted by a power law.
However,we observe
a clear power law decay of persistence for $p>0.655$. 
We can even observe finite size scaling of persistence
at $p=p_s$,
For $p_s=0.655$, we show the asymptotic value of persistence
at various system sizes N. 
We observe a power-law decay of $P_l(\infty)$ as a function
of $N$ and it saturates for $p<p_s$ in the thermodynamic limit 
(See inset of Fig.\ref{fig:4}). If we expect finite size
scaling, we can postulate that $P_l(\infty)\sim N^{z\theta}$.
We find that $z\theta=3.18$ and $\theta=1.92$ implying that $z=1.656$.  
($z<2$ indicating superdiffusive behavior. For 1-D DP, $z=1.58$.)
Fig.\ref{fig:4} show the scaling plot of $P_s(t)$ with $N^{z\theta}$.
Such scaling is not obtained for other values of $p$.

In general, persistence shows
exponential decay in the active phase.
The power-law decay is observed only at the
critical point.  But our case is different. 
Near $p_c$, the persistence decay is slower than
logarithmic  in active state.  This is because active sites spread
logarithmically in time and and it takes very long for inactive sites
to become active. Thus persistence is dominated by 
inactive sites which did nit become active till that time. 
We denote persistence of type A (CDP) sites with initial state 1 and 0 
by $P_l(A,1,t)$ and $P_l(A,0,t)$ respectively.  We define
$P_l(B,1,t)$
and $P_l(B,0,t)$ in an analogous manner for DP sites of type  B.
$P_l(t)$  is the sum of these four quantities.
The total persistence $P_l(t)$ is found to be dictated by $P_l(A,1,t)$.
The quantity $P_l(A,1,t)$ deviate from the initial
condition very slowly for $p>p_s$.
For $p<p_c$, the evolution effectively
stops as soon the system reaches absorbing state and a finite
value of persistence is expected.
For $p>p_c$ the persistence is expected to go to zero asymptotically.
For $p>p_s$ it shows power law decay. 

\begin{figure}[hbt!]
\scalebox{0.3}{
      \includegraphics{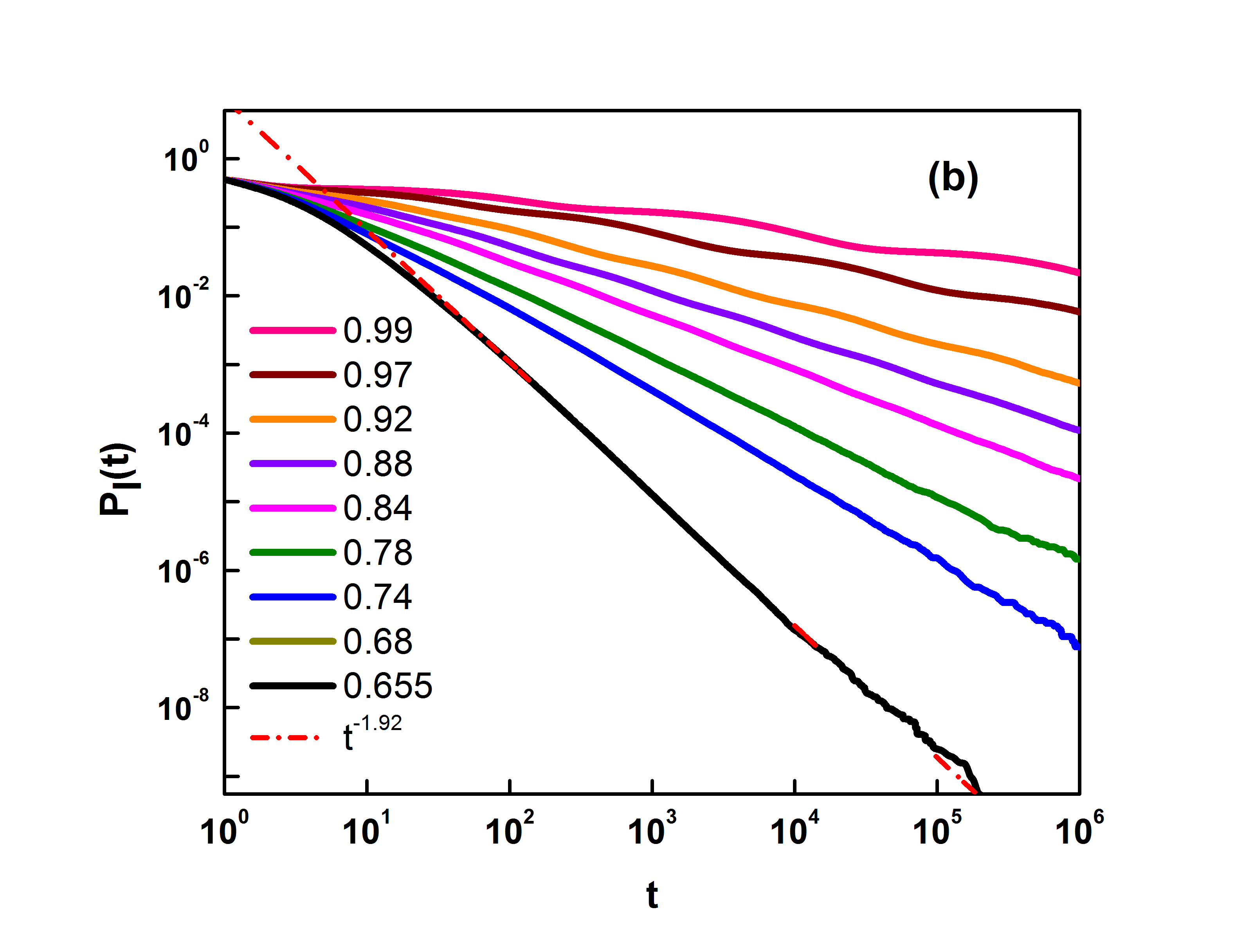}
}
\scalebox{0.3}{
      \includegraphics{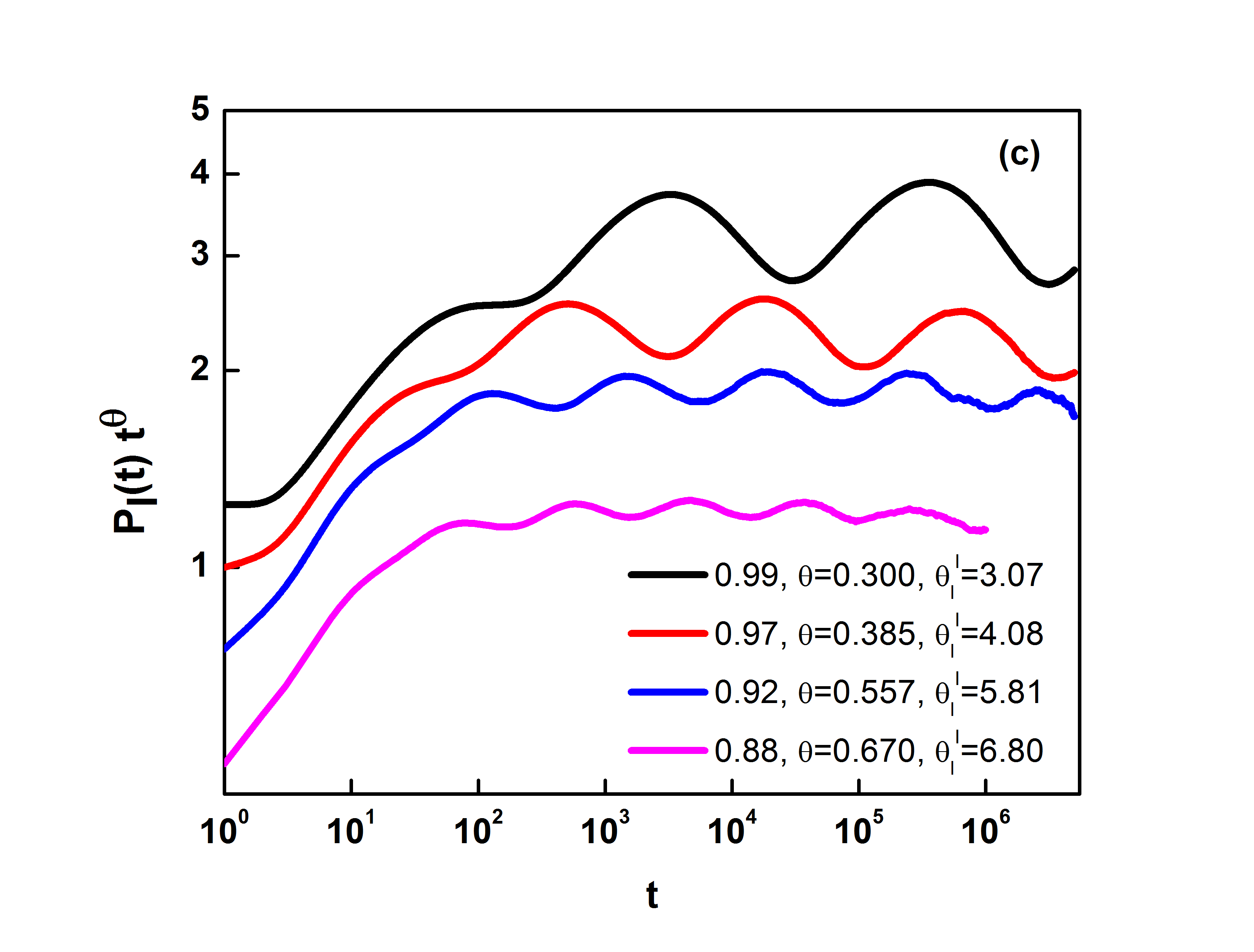}
}
    \caption{ 
	(a) Time evolution of $P_l(t)$ with $r=0.5$ and $p \ge p_s$.
	Clearly the persistence exponent $\theta$ changes
	continuously for $p>p_s$.Power-law is obtained at $p=p_s$.
	(b) Log-log plot of $P_l(t) \times t^{\theta}$ with $r=0.5$ and
	$p>p_s, p \rightarrow 1$. The log-periodic oscillations
	are clearly evident in this figure. The y-axis is multiplied
	by arbitrary constants for better visualization.}
\label{fig:6}
\end{figure}

For $p=p_s$ we observe a clear power-law 
decay of persistence in time with exponent 1.92.
For $p>p_s$, the quantity $P_l(t)$  
continues to decay as power-law  as
shown in Fig.\ref{fig:6}(a). In fact, it decays
with a smaller exponent. Thus the rate at
which system loses the memory of initial conditions is fastest
at this  point! Unlike other 
cases where we observe exponential decay of
persistence in the active phase,
the memory of
initial conditions decay
very slowly even in the active phase.
The behavior can be described by 
$P_l(t)\sim t^{-\theta_l}$ for $p\ge p_s$
where $\theta_l$ is known as the persistence exponent. 
Thus $P_l(t) \times  t^{\theta_l}$
is a constant.For larger values
of $p$ close to 1,
there are systematic oscillations over and 
above the power-law decay
and they can be best described by complex persistence 
exponent. The amplitude of oscillations 
increases as $p\rightarrow 1$.

\begin{figure}[hbt!]
\scalebox{0.3}{
      \includegraphics{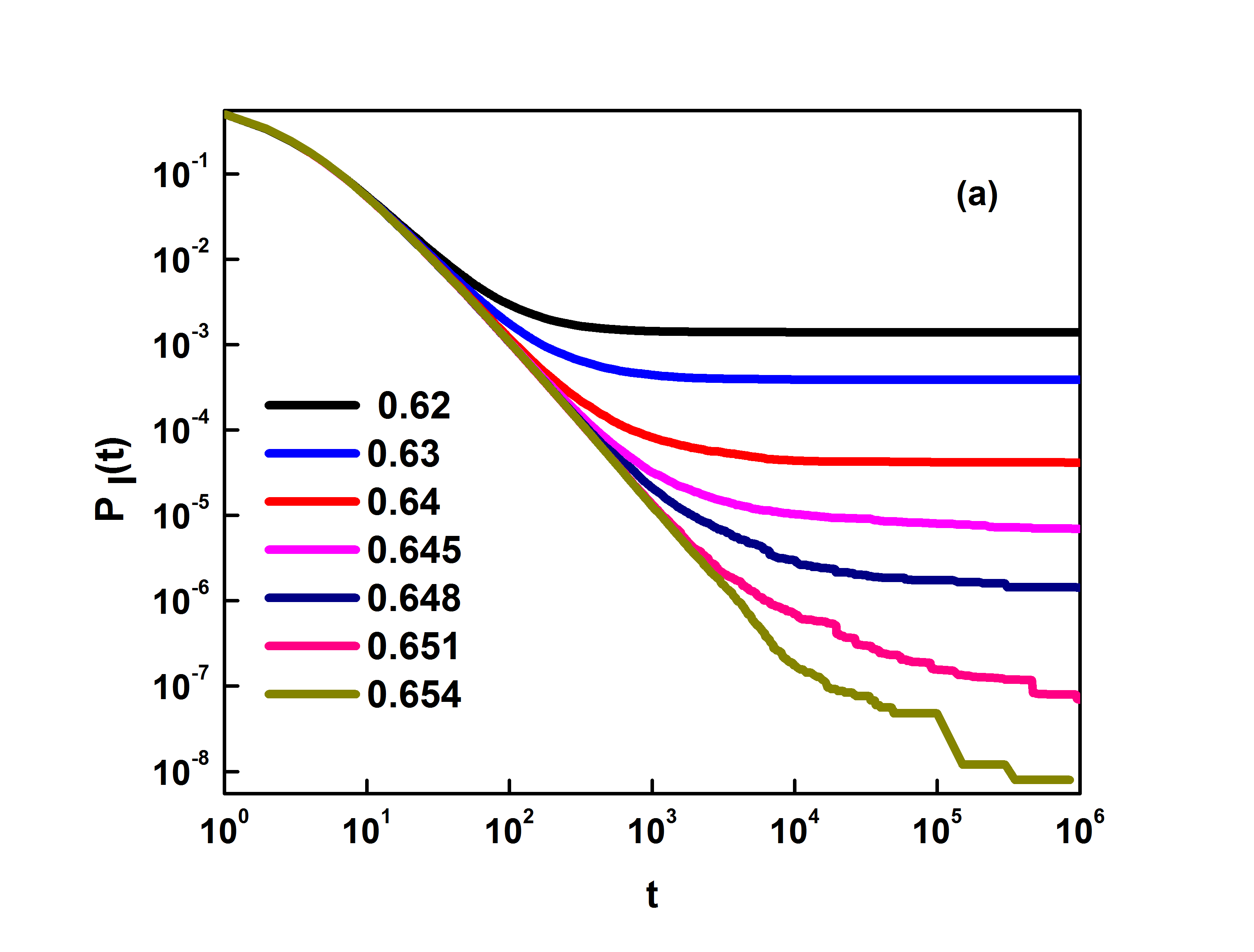}
      }
\caption{ Time evolution of local persistence $P_l(t)$ for a system of
	size $2.5 \times 10^6$ with $r=0.5$ and $p<p_s$.}
\label{fig:5}
\end{figure}

This oscillatory nature of persistence is not reflected in any other
quantity. The number of active sites saturates quickly in few time-steps.
Similarly, rate at which active sites become inactive and \textit{vice-versa}
reaches a constant value quickly. The number of domain walls where active
and inactive sites are next to each other do not show any oscillations.
If we discriminate between different initial conditions and compute
four different quantities, depending on initial condition and whether site
is of DP character or CDP character, we observe that the persistence
is essentially dictated by sites of CDP type which are active initially.
All other persistence goes to zero exponentially fast and logarithmic
oscillations are due to CDP sites which are active in the beginning.
Obviously, the CDP sites form clusters of different sizes, the
probability of size $k$ decreases exponentially with $k$. However,
larger the size it is difficult for 0's to invade the centre of
CDP cluster.

\begin{figure}[hbt!]
\scalebox{0.3}{
      \includegraphics{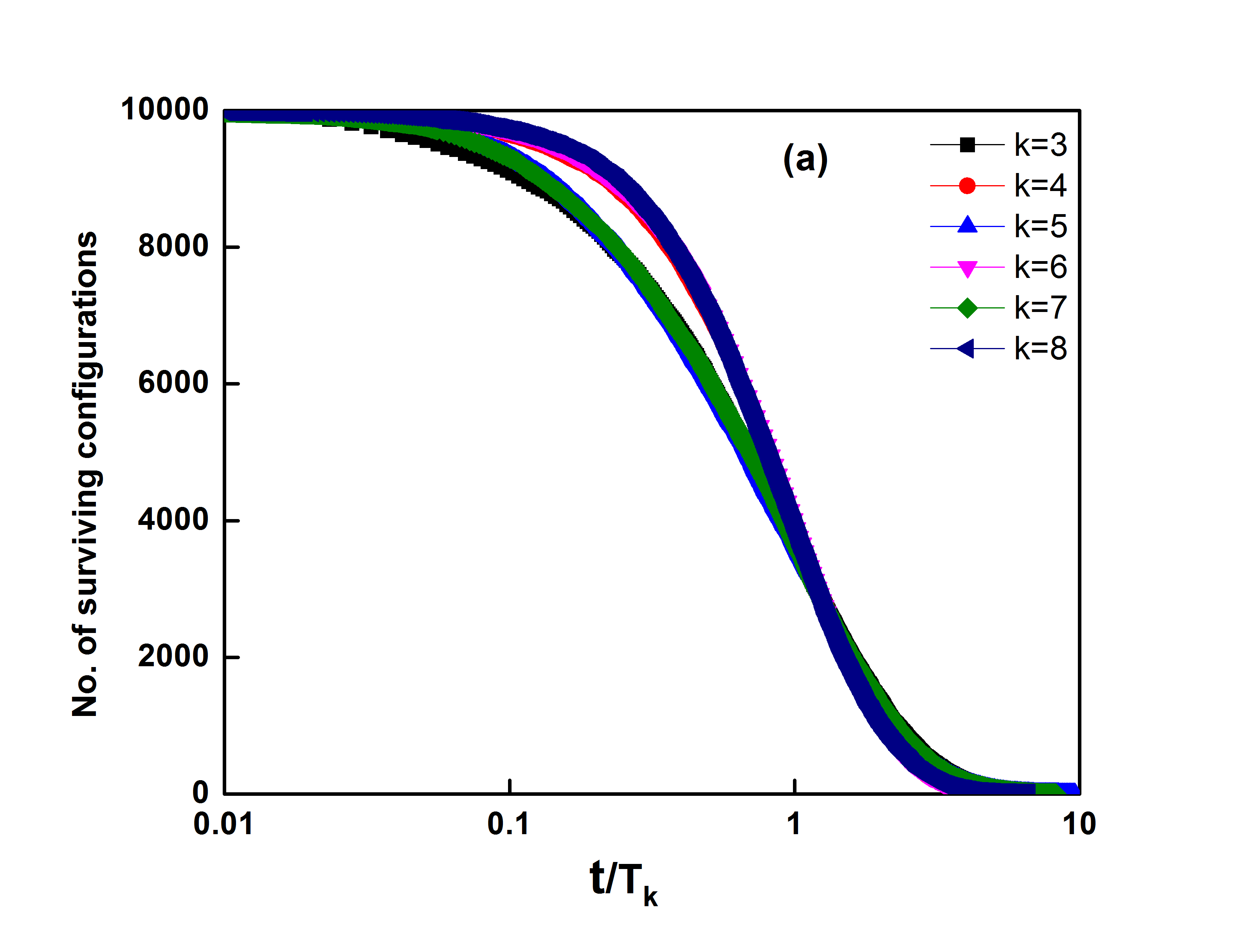}
}
\scalebox{0.3}{
      \includegraphics{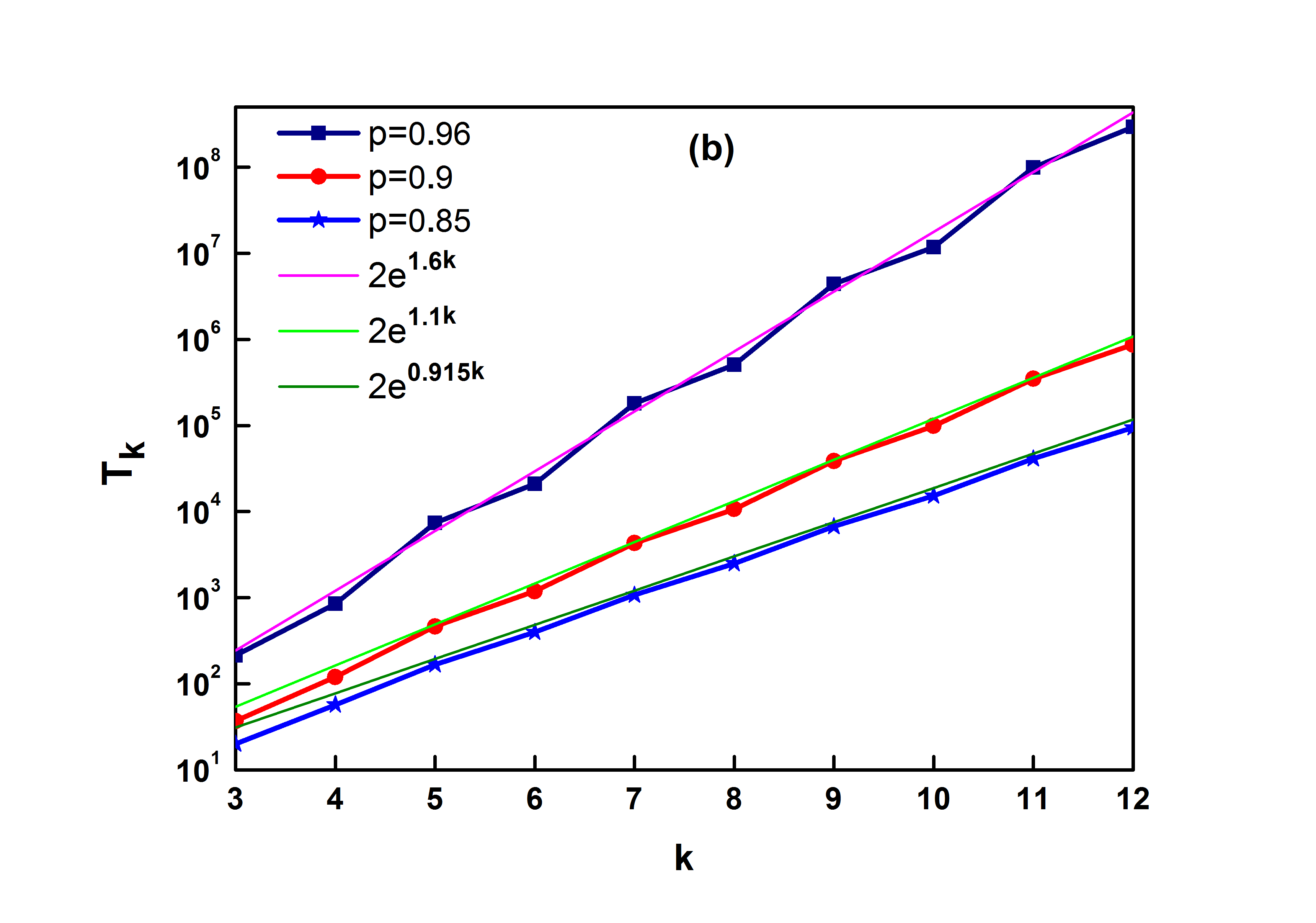}
}
\scalebox{0.3}{
      \includegraphics{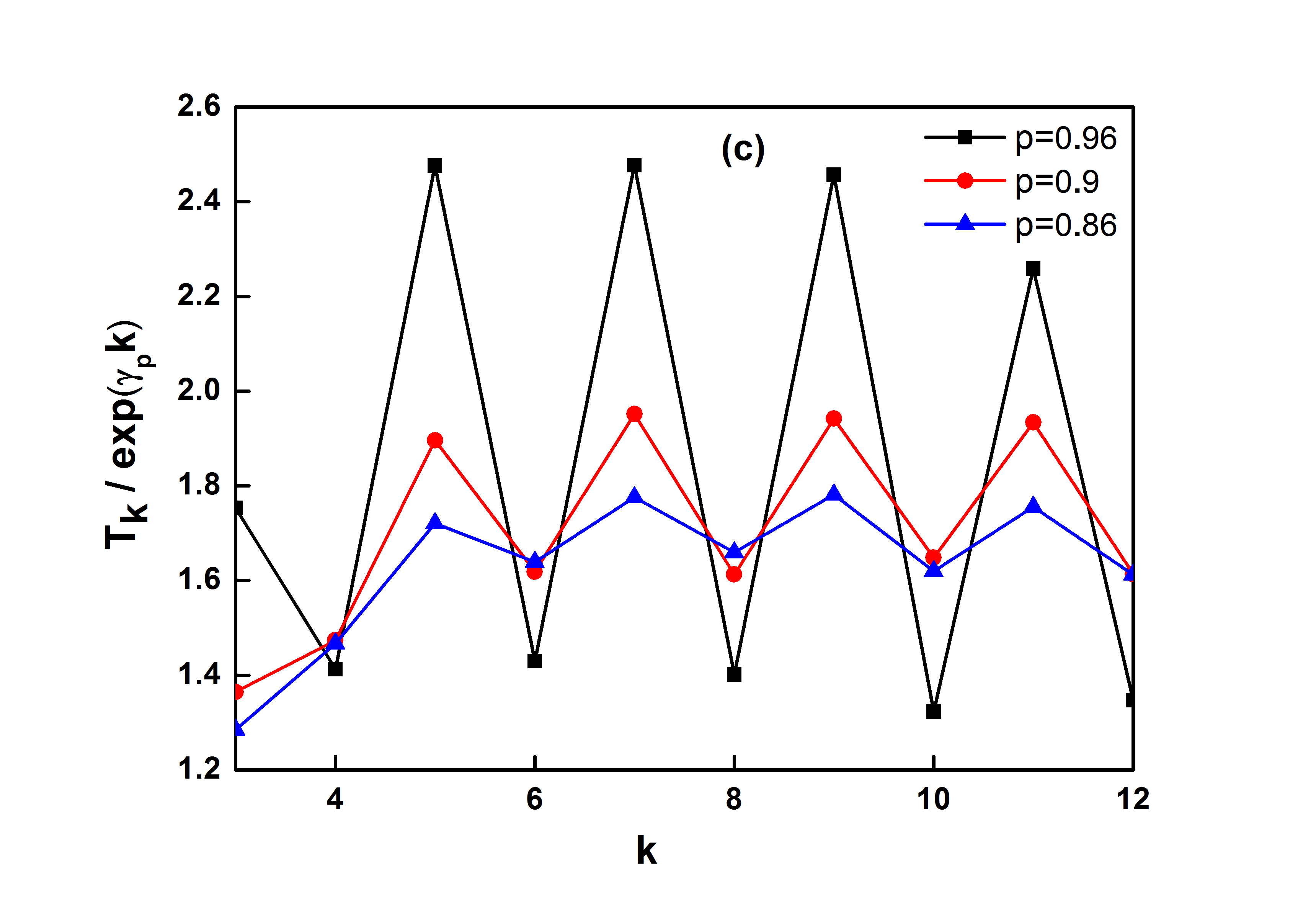}
}
    \caption{(a) We plot the number of surviving configurations
	    $\it{f_s(t)}$ $\it{vs}$ $t/T_k$ for various system of k+2 sites,
	    where k=1,2,4,5,6,7 and $T_k$ is the time taken by k+2
	    sites to become inactive.
	    (b) We plot $<T_k>$ the average time taken by k+2
	    sites to become inactive $\it{vs}$ the number of sites k
	    in the cluster. The plot can be fitted with relationship
	    $exp(\gamma \times k)^2$ where $\gamma=0.915,1.1,1.6$ for
	    $p=0.85,0.9,0.96$.
	    (c) We plot the relation $T_k/exp(\gamma k)$ $\it{vs}$ k. 
}
   \label{fig:7}
\end{figure}

We simulate the systems of $k+2$ sites such that first and $(k+2)^{th}$
site is fixed at 0 and we have a cluster of $k$ sites evolving
according to CDP rules. These sites are active at the beginning. 
If all active sites have become inactive at least once, 
we consider it as a configuration which has not survived.
In Fig.\ref{fig:7}(a), we have plotted the
fraction of surviving configurations
$f_s(t)$ as a function of time.  
We  also  
compute average time by which all these $k$ sites have become inactive
at least once. This time $T_k$ increases exponentially. But there
are oscillations over and above the exponential.
In Fig.\ref{fig:7}(b) we plot $<T_k>$ the average time taken by k+2
sites to become inactive as a function of
number of sites k in cluster.
Fig.\ref{fig:7}(c) shows the plot of the relation $T_k/exp(\gamma k)$
with varying k.
The probability of CDP cluster of size $k$ decreases exponentially.
However, the lifetime of such a cluster  increases exponentially.
The combination of exponentially rare regions which survive
for exponentially long times leads to a power-law and the 
oscillations over and above this exponential lead to log-periodicity.

At longer times, bigger and bigger clusters of CDP sites are
invaded fully and the sites become inactive at least once. There
is certain time-scale at which say, cluster of four sites is invaded and
after a certain time cluster of size five is fully invaded. However, 
apart from exponential increase in time-scales, there is a odd-even
oscillation which could be a reason for logarithmic oscillations
in persistence.

A complex exponent would imply that $\theta_l=\theta+i\theta'$.
Now the behavior is 
given by  $P_l(t)\sim Re(
A t^{-\theta -i\theta'}) \sim A t^{-\theta} \cos(\theta'\log(t))$
and  $P_l(t) t^{\theta} \sim A \cos(\theta'\log(t))$. If we plot
$P_l(t) \times t^{\theta}$ as a function of $t$, we should observe
log-periodic oscillations over and above the constant.
This is precisely the behavior for large values of $p$ as shown in
Fig.\ref{fig:6}(b).
Since the function is log-periodic, it is very
difficult to find exact time-periodicity.
The period of these oscillations
decreases and amplitude increases as $p \rightarrow 1$.
For $p$ close to $p_s$,  the amplitude (if any) is very small. 
and it is difficult to determine
if $\theta' \ne 0$ close to $p_s$. 
Log-periodic oscillations emerge
due to the inherent self-similar structure in a variety of 
studies \cite{newman1995log,akkermans2012spatial}.
In our model the self-similarity is absent.
The value of $\theta$ decreases as we approach $p=1$.
For $p\rightarrow 1$, $\theta \rightarrow 0$.

\begin{figure}
\scalebox{0.3}{
               \includegraphics{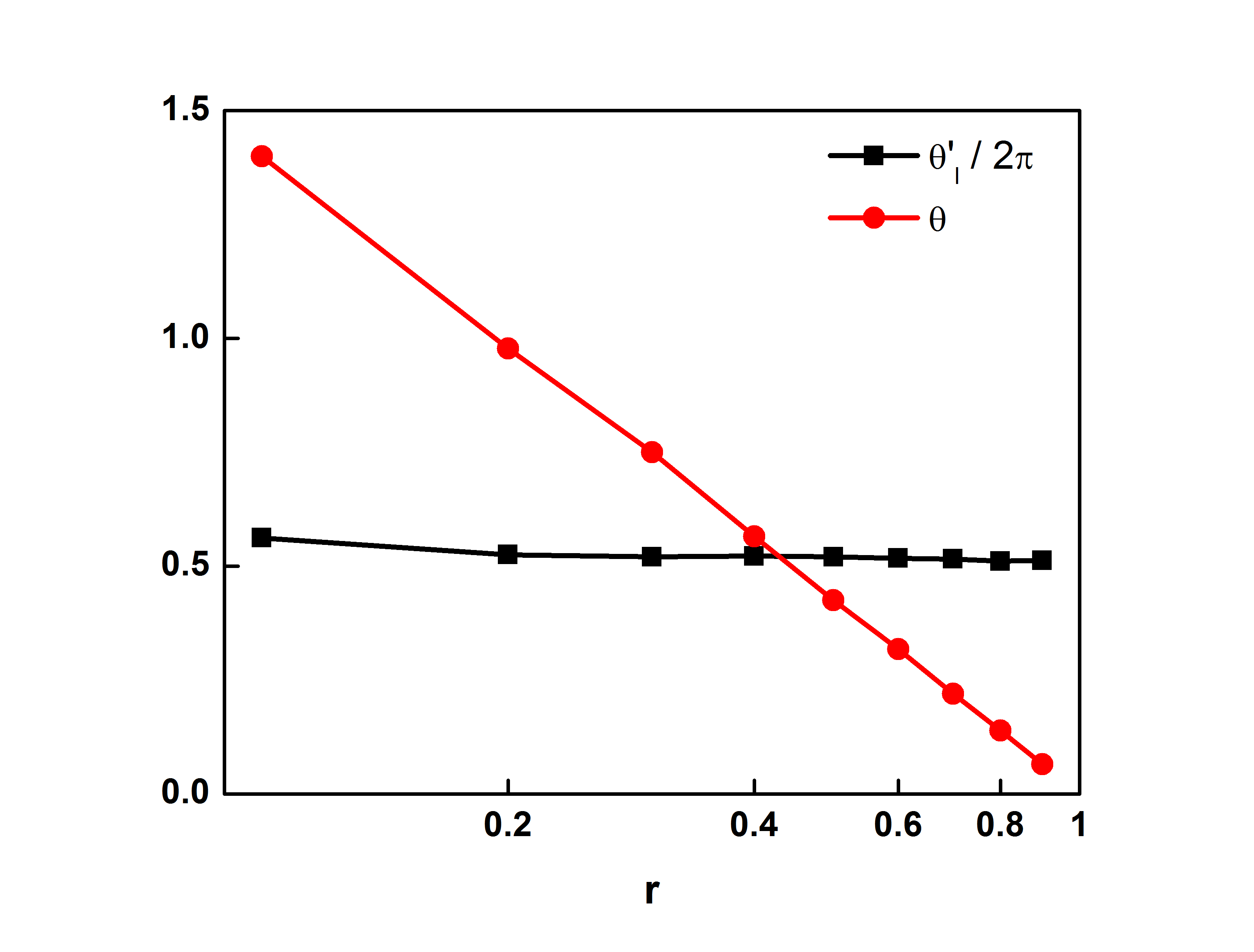}
}
   \caption{\label{fig:8} Semi-logarithmic plot of $Re(\theta_l)=\theta$
	and $Im(\theta_l)/2\pi=\theta'/2\pi$ with $p=0.96$ and varying
	disorder fraction $r$.}
\end{figure}

For CDP, $P_l(\infty)>0$ for any $p$. 
For DP, $P_l(t)$ will decay exponentially
for $p>p_c$. When both types of evolution are possible,
naively one may expect
that the decay will be slower than exponential due to clusters
of CDP sites. One could expect a stretched exponential,
power-law or even logarithmic decay. 
The dynamics will further slow down with an increase of $r$,
leading to a decrease in the real part of the exponent.
Fig.\ref{fig:8} shows that the real part of
persistence exponent varies as log($r$).
Therefore, $\theta\rightarrow 0$ as
$r\rightarrow 1$. However, the
imaginary part of persistence exponent $\theta'$
decreases only slightly with increase in defects $r$.

\section{Summary}
We studied contact
process when a fraction $r$ of sites on 1-D
lattice follows CDP rules, rest evolve according to rules leading
to DP universality class. 
For $r=0.5$, we observe a
transition to the fluctuating phase at critical probability $p_c=0.651$ 
for $r=0.5$. In the absorbing phase, we
observe the Griffiths phase over a range of parameters, 
where the order parameter $\rho(t)$ decays as a 
power-law with continuously varying exponent.
For $p<<p_c$, $\rho(t)$ decays in stretched exponential manner
and eventually shows exponential behavior.
For  $p>p_c$, $\rho(\infty)>0$.
The slow dynamics in the absorbing phase is due to the rare region effect.
The rare region effect decomposes lattice into several
disconnected finite-size clusters. These clusters
are active while the bulk is in the inactive phase.
Thus the overall activity is the sum of
activities of clusters of various sizes.
At $p=p_c$, the decay is extremely slow and we observe a logarithmic decay. 
We confirmed the critical point by large lattice as well as single seed simulation.
We also obtained the survival probability $P_s(t)$ and the average number of active sites
in a cluster starting with single seed $N(t)$. 

We also study the local persistence $P_l(t)$ in this system. 
In DP, critical point $p_s$ at which $P_l(\infty)=0$ coincides
with $p_c$ in general.
For CDP, the persistence $P_l(t)$ does not goes to zero in either phase.
In our case, decay of persistence is slower than power law at $p=p_c$.
For $p>p_s$, $P_l(t)$ decays as power-law 
with continuously varying complex exponent.
The real and imaginary part of the exponent decrease as
$p\rightarrow 1$, but amplitude increases. 
This system does not have a self-similar or fractal
disorder and the underlying lattice is not fractal. 
Interestingly, log-periodic oscillations can be observed
due to uncorrelated quenched disorder alone. 
We have mainly presented results for $r=0.5$.
However, changing $r$ leads to similar
results. The above observations may be applicable to other dynamical rules and
topologies.

\section{Acknowledgement}
PMG thanks DST-SERB for financial assistance and Prof. P. Sen 
and Prof. M. Burma for discussions.

\bibliography{new}

\end{document}